\documentclass[letterpaper,10pt]{article}

\usepackage{amssymb}
\usepackage{amsmath}

\newcommand{\dd}{\textrm{d}}

\def \pre{Preprint}

\author{A. L\'opez-Ortega\thanks{alopezo@ipn.mx} \\
Centro de Investigaci\'on en Ciencia Aplicada y Tecnolog\'{\i}a Avanzada. \\ 
Unidad Legaria. Instituto Polit\'ecnico Nacional. \\
Calzada Legaria \# 694. Colonia Irrigaci\'on. Delegaci\'on Miguel Hidalgo. \\
M\'exico, D.\ F., M\'exico. \\
C.\ P.\  11500  
}

\title{Area spectrum of the $d$-dimensional Reissner-Nordstr\"om black hole in the small charge limit}

\begin{document}

\maketitle

\begin{abstract}

A conjecture by Hod states that for the black hole horizon the spacing of its area spectrum is determined by the asymptotic value of its quasinormal frequencies. Recently to overcome some difficulties, Maggiore proposes some changes to the original Hod's conjecture. Taking into account the modifications proposed by Maggiore we calculate the area quantum of the $d$-dimensional Reissner-Nordstr\"om black hole in the small charge limit.

\end{abstract}



\section{Introduction}
\label{section 1}

Taking into account semiclassical arguments and assuming that the horizon area of a non-extremal black hole behaves as an adiabatic invariant, Bekenstein proposes a discrete and evenly spaced spectrum for the horizon area \cite{Bekenstein:1974jk}--\cite{Bekenstein:1998aw}. Thus, at least in the semiclassical limit, he proposes that the mathematical form of the area spectrum is 
\begin{equation} \label{eq: area spectrum}
 A_n = \epsilon \hbar n, 
\end{equation} 
where $n=0,1,2,\dots$, $\hbar$ stands for the reduced Planck constant, and $\epsilon$ is a dimensionless parameter of order 1. It is believed that a quantum theory of gravity allows us to determine the value of the parameter $\epsilon$ (in the case that the quantum theory confirms that the area spectrum of the black hole horizon takes the form (\ref{eq: area spectrum})).

At present time we do not know a complete quantum theory of gravity. Nevertheless, supposing that the area spectrum is of the form (\ref{eq: area spectrum}) and using different semiclassical methods we can calculate the parameter $\epsilon$ (see \cite{Bekenstein:1995ju}--\cite{Wei:2010yx} for some examples). In the previous references, for the parameter $\epsilon$ often are found the values $\epsilon = 8 \pi$ and $\epsilon = 4 \ln (j)$, with $j=2,3,\dots$. 

We note that several methods used to calculate the area spectrum produce the value $\epsilon = 8 \pi$ \cite{Bekenstein:1974jk,Bekenstein:1973ur,Barvinsky:1996hr,Medved:2009nj}, \cite{Ropotenko:2009mh}--\cite{Medved:2008iq} and for black hole horizons Medved proposes that this value for $\epsilon$ is universal \cite{Medved:2004mh}. Nevertheless, assuming a strict statistical interpretation of the black hole entropy, Bekenstein and  Mukhanov find that the parameter $\epsilon$ must be equal to $\epsilon = 4 \ln (j)$ \cite{Bekenstein:1995ju,Mukhanov:1986me}. In an interesting proposal Hod uses the Bohr correspondence principle and the real part of the asymptotic quasinormal frequencies (QNF) to determine the value of the integer $j$ \cite{Hod:1998vk}. Briefly, Hod proposes that for the four-dimensional Schwarzschild black hole the emission of a quantum produces a change in the black hole mass given by 
\begin{equation}
 \Delta M = \hbar \omega_R,
\end{equation} 
where $\omega_R$ stands for the  real part of the asymptotic QNF.\footnote{This proposal is usually known as Hod's conjecture.} From the known value of $\omega_R$ for the four-dimensional Schwarzschild black hole \cite{Motl:2002hd}--\cite{Andersson:2003fh} Hod finds $j=3$ \cite{Hod:1998vk}. Based on these ideas, Kunstatter expounds a different method to get equivalent results for the area quantum to those by Hod \cite{Kunstatter:2002pj}, at least for some spacetimes. For the results on the area spectrum of the Reissner-Nordstr\"om black hole that produces the original Hod's conjecture see \cite{Setare:2003bd}--\cite{Hod:2007uy}.

Although Hod's conjecture works for Schwarzschild black holes some difficulties arise when it is used in other black holes. For example, for the four-dimensional Kerr black hole it is expected a discrete and equally spaced area spectrum \cite{Bekenstein:1974jk}--\cite{Bekenstein:1998aw}, \cite{Gour:2002ga}, however based on Hod's conjecture Setare and Vagenas find a discrete but not equally spaced area spectrum \cite{Setare:2004uu}. Also it is known that the real part of the asymptotic QNF is not universal, that is, the real part of the asymptotic QNF may depend on the black hole type or even on the field type \cite{Motl:2002hd}--\cite{Andersson:2003fh}, \cite{Natario:2004jd,LopezOrtega:2006vn}.  

Supposing that the quasinormal modes of a black hole can be described as the oscillations of a damped oscillator and taking into account Hod's conjecture, Maggiore \cite{Maggiore:2007nq} proposes that in the semiclassical limit the area spectrum of the event horizon is determined by the asymptotic value of the so-called physical frequency defined by
\begin{equation} \label{eq: physical frequency}
 \omega_{p,k} = \sqrt{\omega_{R}^2 + \omega_{I}^2},
\end{equation} 
where $\omega_I$ stands for the imaginary part of the asymptotic QNF. Making this change and using a similar method to that proposed by Hod \cite{Hod:1998vk}, Maggiore finds that for the four-dimensional Schwarzschild black hole  its area spectrum takes the form (\ref{eq: area spectrum}) with $\epsilon = 8 \pi$. This result for $\epsilon$ coincides with the value calculated by other methods \cite{Medved:2009nj}--\cite{Padmanabhan:2003ub}.

The consequences of Maggiore's proposal have been studied in several spacetimes \cite{Wei:2009yj}--\cite{Wei:2010yx}. Based on the ideas by Hod, Kunstatter, and Maggiore, we calculate the area quantum of the $d$-dimensional Reissner-Nordstr\"om black hole in the small charge limit ($d \geq 4$). We follow the methods already used for the slowly rotating four-dimensional Kerr black hole \cite{Vagenas:2008yi,Medved:2008iq}. Thus for the area spectrum of the Reissner-Nordstr\"om black hole our calculation is an update of the results obtained with the original Hod's conjecture \cite{Setare:2003bd}--\cite{Hod:2007uy}. We also extend the result to non extreme Reissner-Nordstr\"om black hole.

This paper is organized as follows. In Section \ref{section 2} we calculate the area quantum of the $d$-dimensional Schwarzschild black hole. Our main objective in Section \ref{section 2} is to explain the methods that we shall use in the remainder of the paper. Using Hod-Maggiore and Kunstatter-Maggiore methods in Section \ref{section 3}  we calculate the area quantum of the $d$-dimensional Reissner-Nordstr\"om black hole in the small charge limit and then with other method we extend the results to the non extremal Reissner-Nordstr\"om black hole. We also compare our results with those already published. Finally in Section \ref{section 4} we summarize the obtained results.

\section{Area spectrum of the Schwarzschild black hole}
\label{section 2}

Using the ideas by Hod, Kunstatter, and Maggiore \cite{Hod:1998vk,Kunstatter:2002pj,Maggiore:2007nq} in the following section we shall calculate the quantum of area for the $d$-dimensional Reissner-Nordstr\"om  black hole  in the small charge limit. For illustrative purposes in this section we expound a similar calculation for the $d$-dimensional Schwarzschild black hole. As far as we know this computation does not appear in the literature, but it follows from the calculation for the four-dimensional Schwarzschild black hole in a straightforward way \cite{Maggiore:2007nq,Wei:2009yj}.

Notice that in this paper to calculate the area spectra we do not use the original Hod's conjecture \cite{Hod:1998vk}. Our main motivation to use Maggiore's proposal are the examples of spacetimes whose QNF are purely imaginary \cite{LopezOrtega:2009ww} and therefore for their horizons Hod's conjecture predicts a continuous area spectrum. Thus our aim is to investigate the predictions for the area spectra of the Schwarzschild and Reissner-Nordstr\"om black holes that produce the modifications proposed by Maggiore \cite{Maggiore:2007nq}.

The $d$-dimensional Schwarzschild black hole whose metric takes the form
\begin{equation} \label{eq: metric S}
 \dd s^2 = -\left( 1 - \frac{2 \mu }{r^{d-3}} \right) \dd t^2 +  \left( 1 - \frac{2 \mu }{r^{d-3}} \right)^{-1} \dd r^2 + r^2 \dd \Sigma_{d-2}^2, 
\end{equation} 
where $\dd \Sigma_{d-2}^2$ stands for the line element of the $(d-2)$-dimensional unit sphere, the parameter $\mu$ is related to the mass $M$ of the Schwarzschild black hole by (see Appendix A of \cite{Natario:2004jd})
\begin{equation} \label{eq: mass S}
 M= \frac{(d - 2) \Omega_{d-2}}{8 \pi} \mu ,
\end{equation} 
with $\Omega_{d-2}$ denoting the area of an unit $(d-2)$-dimensional sphere
\begin{equation} \label{eq: area unit sphere}
 \Omega_{d-2} = \frac{2 \pi^{(d-1)/2}}{\Gamma(\frac{d-1}{2})} . 
\end{equation} 
We also note that for the $d$-dimensional Schwarzschild black hole the area of its event horizon is given by
\begin{equation} \label{eq: area S}
 A = \Omega_{d-2} r_+^{d-2},
\end{equation}  
where $r_+$ is the radius of its event horizon, $r_+ = (2 \mu)^{1/(d-3)}$.

In order to exploit the ideas by Hod, Kunstatter, and Maggiore to calculate the area and entropy quanta of the black hole under study, we need to know exactly the QNF or at least its asymptotic limit \cite{Hod:1998vk,Kunstatter:2002pj,Maggiore:2007nq}. For the $d$-dimensional Schwarzschild black hole the asymptotic QNF of the gravitational perturbations are equal to \cite{Motl:2002hd,Motl:2003cd,Andersson:2003fh,Natario:2004jd,Birmingham:2003rf} 
\begin{equation}  \label{eq: asymptotic QNF S}
 \omega = \frac{d-3}{4 \pi (2 \mu)^{1/(d-3)}} \ln (3) + i  \frac{d-3}{2 (2 \mu)^{1/(d-3)}} \left(k + \frac{1}{2}  \right), \quad k \in \mathbb{N}, \quad  k \to \infty.
\end{equation}  
From the previous formula we get that for the $d$-dimensional Schwarzschild black hole the physical frequency (\ref{eq: physical frequency}) is
\begin{equation}
 \omega_{p,k} = \frac{d-3}{2 (2 \mu)^{1/(d-3)} }k.
\end{equation} 

We shall use two methods to calculate the area quantum from the asymptotic QNF. The first method is based on the ideas by Hod \cite{Hod:1998vk} and Maggiore \cite{Maggiore:2007nq}. This method works as follows. Based on Hod's ideas, Maggiore proposes that the mass of the four-dimensional Schwarzschild black hole changes in discrete steps determined by \cite{Maggiore:2007nq}
\begin{equation} \label{eq: Maggiore S}
 \Delta M = \hbar \Delta \omega,
\end{equation} 
where $ \Delta \omega$ is equal to
\begin{equation}\label{eq: Delta omega S}
 \Delta \omega = \omega_{p,k+1} -  \omega_{p,k} = \frac{d-3}{2 (2 \mu)^{1/(d-3)}},
\end{equation} 
with $d=4$. We propose that formula (\ref{eq: Maggiore S}) is valid for the $d$-dimensional Schwarzschild black hole  and taking into account formulas (\ref{eq: mass S}), (\ref{eq: area unit sphere}), and (\ref{eq: area S}) we find that a small change in the mass of the Schwarzschild black hole produces a change in its horizon area given by
\begin{equation} \label{eq: delta A S}
 \Delta A = \frac{2 (2 \mu)^{1/(d-3)}}{d-3} 8 \pi \Delta M. 
\end{equation} 

Using formulas (\ref{eq: Maggiore S}) and (\ref{eq: Delta omega S}) we find that the area quantum of the $d$-dimensional Schwarzschild black hole is 
\begin{equation} \label{eq: area quantum S}
 \Delta A = 8 \pi \hbar.
\end{equation} 
Thus $\epsilon = 8 \pi$ for this black hole and from the Bekenstein-Hawking area-entropy relation \cite{Wald:1999vt} we obtain the entropy quantum 
\begin{equation} \label{eq: entropy quantum S}
 \Delta S = 2 \pi .
\end{equation} 

The second method that we shall use to calculate the area quantum of the event horizon is based on the ideas by Kunstatter \cite{Kunstatter:2002pj} and Maggiore \cite{Maggiore:2007nq}. This method works as follows. First we note that for a system with energy $E$ and oscillation frequency $\omega$ the quantity 
\begin{equation}
 I = \int \frac{\dd E}{\omega}
\end{equation} 
is an adiabatic invariant \cite{Kunstatter:2002pj}. Furthermore we recall that in the semiclassical limit Bohr-Sommerfeld quantization rule states that the adiabatic invariant $I$ has an equally spaced spectrum, thus $I  =  n \hbar$, $n \in \mathbb{N}$ \cite{Kunstatter:2002pj}. From these facts and taking into account the ideas by Kunstatter \cite{Kunstatter:2002pj} and Maggiore \cite{Maggiore:2007nq}, it is proposed that for the calculation of the area spectrum the quantity $\Delta \omega$ of formula (\ref{eq: Delta omega S}) is the appropriate oscillation frequency \cite{Vagenas:2008yi}. Thus for the Schwarzschild black hole we need to calculate the adiabatic invariant \cite{Kunstatter:2002pj,Wei:2009yj,Vagenas:2008yi}
\begin{equation}  \label{eq: adiabatic invariant S}
 I = \int \frac{\dd M}{\Delta \omega}.
\end{equation} 

For the $d$-dimensional Schwarzschild black hole we get 
\begin{eqnarray}
I &= \frac{d-2}{d-3}\frac{2^{(d-2)/(d-3)} \Omega_{d-2}}{8 \pi} \int \mu^{1/(d-3)} \dd \mu = \frac{A}{8 \pi}. 
\end{eqnarray} 
Hence, Bohr-Sommerfeld quantization rule and the previous expression for $I$ imply that in the semiclassical limit the area spectrum of the  $d$-dimensional Schwarzschild black hole takes the form
\begin{equation} \label{eq: area spectrum S}
 A_n = 8 \pi \hbar n .
\end{equation} 
From this expression for the area spectrum we find the area quantum $\Delta A$ of formula (\ref{eq: area quantum S}) (and therefore the entropy quantum $\Delta S$ of expression (\ref{eq: entropy quantum S})). Thus for the Schwarzschild black hole the two methods produce identical values for $\Delta A$ and $\Delta S$. 

We notice that expression (\ref{eq: area quantum S}) for the area quantum and formula (\ref{eq: entropy quantum S}) for the entropy quantum of the $d$-dimensional Schwarzschild black hole are equal to those previously obtained for the four-dimensional Schwarzschild black hole \cite{Maggiore:2007nq}, \cite{Wei:2009yj}. Thus for the Schwarzschild black hole the area and entropy quanta are independent of the spacetime dimension.

At this point we note that in the method based on Hod \cite{Hod:1998vk} and Maggiore \cite{Maggiore:2007nq} proposals the area spectrum of the event horizon has the mathematical form (\ref{eq: area spectrum}), that is, a discrete and equally spaced area spectrum is assumed and the method gives us the value of the area quantum. In contrast the method based on ideas by Kunstatter \cite{Kunstatter:2002pj} and Maggiore \cite{Maggiore:2007nq} produces a mathematical expression for the area spectrum and the value for the area quantum is a by product.

\section{Area spectrum of the Reissner-Nordstr\"om black hole}
\label{section 3}

In this section we use Hod's conjecture \cite{Hod:1998vk} with the changes suggested by Maggiore \cite{Maggiore:2007nq} to calculate the area quantum of the $d$-dimensional Reissner-Nordstr\"om black hole in the small charge limit and then with a different method we extend these results to non extreme Reissner-Nordstr\"om black hole. Hence for the Reissner-Nordstr\"om black hole we revise and update some results obtained with the original Hod's proposal \cite{Setare:2003bd}--\cite{Hod:2007uy}. Our calculation for the $d$-dimensional Reissner-Nordstr\"om black hole in the small charge limit is similar to that for the slowly rotating four-dimensional Kerr black hole \cite{Vagenas:2008yi,Medved:2008iq}.

The $d$-dimensional Reissner-Nordstr\"om black hole whose metric is 
\begin{equation} \label{eq: metric RN}
 \dd s^2 = -\left( 1 - \frac{2 \mu }{r^{d-3}} + \frac{q^2}{r^{2(d-3)} } \right) \dd t^2 +  \left( 1 - \frac{2 \mu }{r^{d-3}} + \frac{q^2}{r^{2(d-3)}} \right)^{-1} \dd r^2 + r^2 \dd \Sigma_{d-2}^2, 
\end{equation} 
where we define $\dd \Sigma_{d-2}^2$ as in the previous section, the parameters $\mu$ and $q$ are related to the ADM mass $M$ and electric charge $Q$ of the black hole by the expressions \cite{Natario:2004jd}
\begin{equation} \label{eq: mass charge RN}
 \mu = \frac{8 \pi }{\Omega_{d-2} (d-2)} M, \qquad q^2 = \frac{2}{(d-2)(d-3)}Q^2,
\end{equation} 
with $\Omega_{d-2}$ already defined in formula (\ref{eq: area unit sphere}).

For the  $d$-dimensional Reissner-Nordstr\"om black hole the area of its event horizon $A$ and its electrostatic potential at the horizon $\Phi_+$  are
\begin{equation} \label{eq: area potential RN}
 A = \Omega_{d-2} r_+^{d-2}, \qquad \Phi_+ = \frac{\Omega_{d-2} Q}{4 \pi (d-3) r_+^{d-3}},
\end{equation} 
where $r_+$ stands for the radius of the event horizon
\begin{equation}
 r_+^{d-3} = \mu + \sqrt{\mu^2 - q^2}.
\end{equation} 
We note that the radius of the inner horizon is determined by
\begin{equation}
 r_-^{d-3} = \mu - \sqrt{\mu^2 - q^2}.
\end{equation} 

First we focus on the area quantum of the $d$-dimensional Reissner-Nordstr\"om black hole in the small charge limit, thus, we assume that $q \ll \mu$. Our main reasons are
\begin{enumerate}
 \item It is believed that extreme and near extreme black holes are highly quantum objects \cite{Barvinsky:2000gf,Medved:2009nj,Gour:2002ga,Gour:2002uk}, therefore we think that the semiclassical methods explained in the previous section are not useful for extreme or near extreme black holes.
 \item In the small charge limit there is an explicit formula for the asymptotic QNF of the Reissner-Nordstr\"om black hole (see formula (3.20) of \cite{Daghigh:2006hu}). Generally the asymptotic QNF of the Reissner-Nordstr\"om black hole are given implicitly \cite{Motl:2003cd,Andersson:2003fh,Natario:2004jd,Daghigh:2006hu}.
\end{enumerate}

For the $d$-dimensional Reissner-Nordstr\"om black hole in the small charge limit the asymptotic QNF of the coupled gravitational and electromagnetic perturbations reduce to \cite{Daghigh:2006hu}
\begin{eqnarray}  \label{eq: asymptotic QNF RN}
 \omega &= \frac{d-3}{4 \pi (2 \mu)^{1/(d-3) } } \ln \left(3 + 4 \cos \left( \frac{d-3}{2d -5} \pi \right) \right)  +   \frac{i (d-3)}{2 (2 \mu)^{1/(d-3)}} \left(k + \frac{1}{2}  \right), 
\end{eqnarray} 
with  $k \in \mathbb{N}$, $k \to \infty$. It is convenient to comment that the asymptotic QNF (\ref{eq: asymptotic QNF RN}) do not correspond to those for the Schwarzschild black hole (\ref{eq: asymptotic QNF S}) as the electric charge goes to zero. In the zero charge limit the difference between (\ref{eq: asymptotic QNF S}) and (\ref{eq: asymptotic QNF RN}) is in the logarithm term of the real part for the asymptotic QNF. For example, when $d=4$ in the result for Schwarzschild appears the factor $\ln(3)$ while in the result for Reissner-Nordstr\"om the corresponding factor is $\ln(5)$. In \cite{Daghigh:2006hu} it is shown that this fact is due to an discrete change in the topology of the Stokes lines used in the computation when the damping of the QNF for the Reissner-Nordstr\"om black hole  increases. 

Furthermore in \cite{Daghigh:2006hu} evidence is presented that for the asymptotic QNF of the Reissner-Nordstr\"om black hole  when we increase the damping there is a range  such that the real part varies from the Schwarzschild value to near zero and then the real part goes to the value for the Reissner-Nordstr\"om black hole given in (\ref{eq: asymptotic QNF RN}). Since we only use the methods with the modifications proposed by Maggiore for our problem this fact does not change the analysis, because we must calculate the physical frequency in the limit $k \to \infty$ and in this case the real part of the asymptotic QNF becomes negligible and hence its variation does not change in an appreciable way the value of the physical frequency. Moreover notice that the imaginary part of the asymptotic QNF is continuous when we go from the asymptotic QNF for the Reissner-Nordstr\"om black hole to those for the Schwarzschild black hole.

From the asymptotic QNF (\ref{eq: asymptotic QNF RN}) we get the quantity $\Delta \omega$ corresponding to the Reissner-Nordstr\"om black hole in the small charge limit is
\begin{equation}\label{eq: Delta omega RN}
 \Delta \omega = \frac{d-3}{2 (2 \mu)^{1/(d-3)}} ,
\end{equation} 
and as for the $d$-dimensional Schwarzschild black hole we propose that formula (\ref{eq: Maggiore S}) is valid for the $d$-dimensional Reissner-Nordstr\"om black hole (for the four-dimensional Kerr black hole see \cite{Vagenas:2008yi,Medved:2008iq}). Hence from formulas (\ref{eq: Maggiore S}) and (\ref{eq: mass charge RN}) we find
\begin{equation}\label{eq: Delta M RN}
 \Delta M = \frac{\hbar (d - 3)}{2 (2 \mu )^{1/(d-3)}} .
\end{equation} 

Taking into account the first formula of (\ref{eq: area potential RN}) we obtain that, in the small charge limit, a small variation in the mass produces a change in the horizon area of the Reissner-Nordstr\"om black hole equal to
\begin{eqnarray} \label{eq: RN Delta A}
 \Delta A &\approx \Omega_{d-2} \frac{d-2}{d-3} (2 \mu)^{1/(d-3)} 2 \Delta \mu  \approx \frac{2 (2\mu)^{1/(d-3)}}{d-3} 8 \pi \Delta M ,
\end{eqnarray} 
and using formula (\ref{eq: Delta M RN}) we get that the area quantum of the $d$-dimensional Reissner-Nordstr\"om black hole in the small charge limit is $\Delta A = 8 \pi \hbar$ (and hence $\epsilon = 8 \pi$).

Also, to calculate the area quantum of the  Reissner-Nordstr\"om black hole in the small charge limit we can use the second method explained in the previous section. To leading order, that is, keeping terms of order $(q/2 \mu)^2$ in the calculation, we note that in the small charge limit the area of the Reissner-Nordstr\"om event horizon is approximately
\begin{equation} \label{eq: area RN limit}
 A \approx \Omega_{d-2} (2 \mu)^{(d-2)/(d-3)} - \Omega_{d-2} \frac{d-2}{d-3} \frac{q^2}{(2 \mu)^{(d-4)/(d-3)}}.
\end{equation} 

Following Vagenas \cite{Vagenas:2008yi} and Medved \cite{Medved:2008iq} we propose that for the Reissner-Nordstr\"om black hole the adiabatic invariant $I$ of expression (\ref{eq: adiabatic invariant S}) transforms into
\begin{equation} \label{eq: adiabatic invariant RN}
 I = \int \frac{\dd M - \Phi_+ \dd Q}{\Delta \omega},
\end{equation} 
(see also \cite{Setare:2004uu,LopezOrtega:2009ww,Wei:2010yx}). To leading order we find that for the Reissner-Nordstr\"om black hole in the small charge limit the adiabatic invariant (\ref{eq: adiabatic invariant RN}) becomes
\begin{eqnarray}\label{eq: I RN}
 I & = \frac{(d-2) \Omega_{d-2}}{4 \pi (d-3)} \int (2 \mu)^{1/(d-3)} \dd \mu - \frac{\Omega_{d-2} 2^{(d-2)/(d-3)}}{8 \pi (d-3)^2 \mu^{(d-4)/(d-3)}} \int Q \dd Q \nonumber\\
& \approx \frac{\Omega_{d-2}}{8 \pi}  (2 \mu)^{(d-2)/(d-3)}   - \frac{\Omega_{d-2}}{8 \pi}  \frac{d-2}{d-3} \frac{q^2}{(2 \mu)^{(d-4)/(d-3)}}.
\end{eqnarray} 

Comparing formulas (\ref{eq: area RN limit}) and (\ref{eq: I RN}) we find that in the small charge limit 
\begin{equation} \label{eq: I A RN}
 I \approx \frac{A}{8 \pi},
\end{equation} 
and from Bohr-Sommerfeld quantization rule we get that  for the $d$-dimensional Reissner-Nordstr\"om black hole in the small charge limit its area spectrum to leading order is $A_n \approx 8 \pi \hbar n$. Therefore, to leading order, the area quantum that we get with the second method is identical to that previously calculated.

As we comment at the end of Section 2, the first method that we use previously gives the value for the area quantum, while the second method produces an expression for the area spectrum. Doubtless, in the framework of the Kunstatter method, it is convenient to find the expression of the area spectrum when we kept higher order terms in the expansion in $(q / 2 \mu)$ already used. Keeping terms of order $(q/2 \mu)^4$, we find that for the Reissner-Nordstr\"om black hole in the small charge limit the area of its event horizon is (compare with (\ref{eq: area RN limit}))  
\begin{eqnarray} \label{eq: area RN limit 2}
 A &\approx \Omega_{d-2} (2 \mu)^{(d-2)/(d-3)} - \Omega_{d-2} \frac{d-2}{d-3} \frac{q^2}{(2 \mu)^{(d-4)/(d-3)}}  \nonumber \\
& \quad -\frac{\Omega_{d-2}}{2}  \frac{d-2}{d-3} \frac{2d-7}{d-3} \frac{q^4}{(2 \mu)^{(3d - 10)/(d-3)}} .
\end{eqnarray} 
In a similar way, to the same order in $(q/2 \mu)$, we get that the adiabatic invariant $I$ (\ref{eq: adiabatic invariant RN}) is
\begin{eqnarray}\label{eq: I RN 2}
I  & \approx \frac{1}{8 \pi}  \Omega_{d-2}  (2 \mu)^{(d-2)/(d-3)}   - \frac{1}{8 \pi} \Omega_{d-2} \frac{d-2}{d-3} \frac{q^2}{(2 \mu)^{(d-4)/(d-3)}} \nonumber \\
& \qquad - \frac{1}{8 \pi}  \frac{\Omega_{d-2}}{2}  \frac{d-2}{d-3} \frac{q^4}{(2 \mu)^{(3d - 10)/(d-3)}} .
\end{eqnarray} 

From these expressions we obtain that to this order the area of the event horizon $A$ and the adiabatic invariant $I$ do not satisfy (\ref{eq: I A RN}). Thus to higher order and with the approximations that we use to find expression (\ref{eq: I RN 2}), we get that Kunstatter-Maggiore method predicts that for the Reissner-Nordstr\"om black hole in the small charge limit its area spectrum does not grow linearly. Hence to leading order it is true that both methods give the same area quantum, but to higher order, Kunstatter method does not predict an equally space area spectrum and therefore the area quantum that  it gives is different from that produces the first method.\footnote{I thank to Referee for suggesting this analysis.} For the already studied slowly rotating four dimensional Kerr black hole \cite{Vagenas:2008yi}, \cite{Medved:2008iq}, \cite{Myung:2010af} we guess that only to leading order the two methods used to calculate the area quantum produce the same value. 

We note that in the calculation of the expressions (\ref{eq: area RN limit 2}) and (\ref{eq: I RN 2}) for the area and the quantity $I$ we expand to order $(q/2\mu)^4$ the factor $(1-q^2/\mu^2)^{1/2}$, but we use the physical frequency given in formula (\ref{eq: Delta omega RN}). It is convenient to calculate the corrections of the asymptotic QNF for the Reissner-Nordstr\"om black hole, and use these to find a new expression for the adiabatic invariant $I$, in order to determine whether it satisfies relation (\ref{eq: I A RN}). 

Thus, to leading order, for the $d$-dimensional Reissner-Nordstr\"om black hole in the small charge limit its area quantum is equal to that of the $d$-dimensional Schwarzschild black hole. As for the Schwarzschild black hole, to leading order the Hod-Maggiore and Kunstatter-Maggiore  methods yield that the area quantum of the Reissner-Nordstr\"om black hole in the small charge limit does not depend on the spacetime dimension. Moreover, for these two black holes the two methods that we have used to calculate their area spectra to leading order produce identical results.

From our results we point out that for the Schwarzschild black hole and the Reissner-Nordstr\"om black hole in the small charge limit, the area and entropy spectra are equally spaced to leading order, but we point out that there are black holes for which their entropy spectra are evenly spaced but not their area spectra \cite{Wei:2009yj,Kothawala:2008in}.

Notice that when we use the first method (see formula (\ref{eq: RN Delta A})) we implicitly assume that the change in the horizon area is determined by the change in the mass of the black hole, and thus we ignore the changes in the other parameters (and the changes that these produce in the horizon area). In particular for the Reissner-Nordstr\"om black hole we assume that the electric charge does not change or its change is much smaller than that of the mass. 

Recently Kwon and Nam propose that the appropriate adiabatic invariant to be quantized is similar to that of the Schwarzschild black hole (see formula (\ref{eq: adiabatic invariant S})), even for charged and rotating black holes, in contrast to the proposals by Vagenas \cite{Vagenas:2008yi}, Medved \cite{Medved:2008iq}, and of formula (\ref{eq: adiabatic invariant RN}). For the three-dimensional spinning BTZ black hole they find  that the addition and subtraction of the areas for the outer and inner horizons are quantized. Using the method described in \cite{Kwon:2010um},  Kwon and Nam \cite{Kwon:2010mt} analyze the case of the Reissner-Nordstr\"om black hole and they obtain a result consistent with that presented above for the area quantum in the small charge limit. 

It is convenient to comment that in \cite{Kwon:2010mt} are quantized the surface areas of the event and inner horizons for the Reissner-Nordstr\"om black hole, similar to the proposal by Makela and Repo \cite{Makela:1997rx} and the results obtained with the original Hod's conjecture \cite{Hod:2005dc}, \cite{Hod:2007uy}. Furthermore we note that in the process of quantization,  Kwon and Nam \cite{Kwon:2010mt} propose that we must consider on equal basis the quasinormal modes, the total transmission modes, and the total reflection modes, but in their calculation of the area spectrum of the Reissner-Nordstr\"om black hole they only consider the total transmission modes and the total reflection modes, thus in their computation they do not take into account the QNM, because for the coupled electromagnetic and gravitational perturbations an explicit expression for the asymptotic QNF is not known (except in the small charge limit).

Nevertheless for the Reissner-Nordstr\"om black hole the asymptotic QNF of the uncharged massless Dirac field can be determined in explicit form. For the four-dimensional Reissner-Nordstr\"om black hole the asymptotic QNF of the Dirac field are calculated by Cho in \cite{Cho:2005yc} (see formula (55) of  \cite{Cho:2005yc}). We note that in the $d$-dimensional Reissner-Nordstr\"om black hole the Dirac equation simplifies to a pair of Schr\"odinger type equations with effective potentials \cite{LopezOrtega:2009qc}
\begin{align} \label{e: effective potentials Dirac}
 V_\pm &= - \kappa^2 \left( \frac{1}{r^2} - \frac{2 \mu}{r^{d-1} } + \frac{q^2}{r^{2d -4}} \right) \nonumber \\
& \mp i \kappa \left( 1 - \frac{2 \mu}{r^{d-3} } + \frac{q^2}{r^{2d -6}} \right)^{1/2} \left( - \frac{1}{r^2}  + \frac{\mu (d-1)}{r^{d-1}} - \frac{q^2(d-2)}{r^{2d -4}} \right),
\end{align} 
where $\kappa$ is related to the eigenvalues of the Dirac operator on the $(d-2)-$dimensional sphere. From expressions (\ref{e: effective potentials Dirac}) for the effective potentials we get
\begin{equation}
 \lim_{r \to \infty} V_\pm = 0, \qquad \qquad  \lim_{r \to 0} V_\pm \approx \pm \frac{\alpha}{x^{1 + (d-2)/(2 d - 5)}} ,
\end{equation} 
with $\alpha$ a constant and $x$ is the tortoise coordinate of the Reissner-Nordstr\"om black hole.

Hence we can use the results of \cite{Motl:2003cd}, \cite{Natario:2004jd} to find that in the $d$-dimensional Reissner-Nordstr\"om black hole the  asymptotic QNF of the Dirac field are determined by formula (3.20) of \cite{Natario:2004jd} with $j=1$.\footnote{This parameter $j$ is different from that used in Section 1.} Thus for $d \geq 4$ the asymptotic QNF of the Dirac field are
\begin{equation} \label{e: Dirac QNF}
 \omega = i \kappa^+ k
\end{equation} 
where
\begin{equation}
 \kappa^+ = \frac{d-3}{2 r_+}\left( 1 - \frac{r_-^{d-3}}{r_+^{d-3}} \right)
\end{equation} 
is the surface gravity of the event horizon for the Reissner-Nordstr\"om black hole. Notice that the asymptotic QNF (\ref{e: Dirac QNF}) are purely imaginary and in contrast to QNF (\ref{eq: asymptotic QNF RN}) these are not restricted to the small charge limit.

From (\ref{e: Dirac QNF}) for the Dirac field we obtain that the step in the physical frequency is $\Delta \omega = \kappa^+$. Following Kwon and Nam \cite{Kwon:2010um}, \cite{Kwon:2010mt},  instead the adiabatic invariant (\ref{eq: adiabatic invariant RN}) in what follows we consider the quantity
\begin{equation} \label{e: Kwon Nam I}
 I_{KN} = \int \frac{\dd M }{\Delta \omega} .
\end{equation} 
Making the change of variable $u = \mu + \sqrt{\mu^2 - q^2}$ we find that the value of the previous integral is
\begin{equation} \label{e: Kwon Nam area I}
 I_{KN} = \frac{\Omega_{d-2} r_+^{d-2} }{8 \pi} = \frac{A}{8 \pi}.
\end{equation} 

According to the proposal by Kwon and Nam in the semiclassical limit $I_{KN}$ has an equally spaced spectrum of the form $I_{KN} = n \hbar $. Hence we obtain that for  the $d$-dimensional Reissner-Nordstr\"om black hole the area quantum of its event horizon is $\Delta A = 8 \pi \hbar$ which is consistent with that already obtained. It is convenient to remark that this result is valid for the non-extremal Reissner-Nordstr\"om black holes and it is not restricted to the small charge limit.

Notice that Kwon and Nam state that $I_{KN}$ (\ref{e: Kwon Nam I})  is an action variable and therefore in the proposal by Kunstatter and Maggiore it is the physical quantity that must be quantized. We believe that for the quantity $I_{KN}$ another interpretation is possible. The quantity $I_{KN}$ is quantized when we assume that the changes in the charge or angular momentum of the black hole are neglected and hence only the changes in the mass are relevant. From this viewpoint the previous result that we obtain for the Dirac field deserves additional study.

Based on his conjecture \cite{Hod:1998vk}, Hod studies the area spectrum of the four-dimensional Reissner-Nordstr\"om black hole \cite{Hod:2005dc,Hod:2007uy}. Taking into account the existence of two distinct families for the asymptotic QNF of the charged Klein-Gordon field (see formulas (5) and (6) of \cite{Hod:2007uy}), he finds that a family of QNF leads to an area quantum equal to $\Delta A = 4 \hbar \ln(2)$ for the surface area of the event horizon, whereas the second family of QNF leads to an area quantum equal to $\Delta A = 4 \hbar \ln(3)$ for the total surface area (the area of the outer horizon plus the surface area of the inner horizon). The last result reminds us the proposal by Makela and Repo \cite{Makela:1997rx}, who suggest that for a multihorizon black hole the object that must be quantized in equal steps is the total area of the horizons.

Notice that, in the small charge limit, formulas (5) and (6) of \cite{Hod:2007uy} lead to the same physical frequency and using the previous methods we get that, in the small charge limit, both families of asymptotic QNF lead to the same result for the area quantum of the Reissner-Nordstr\"om event horizon in four dimensions.

It is convenient to comment that for the Reissner-Nordstr\"om black hole with $d \geq 4$, the additional dimensions are included in the term $\dd \Sigma^2_{d-2}$ of the metric (\ref{eq: metric RN}) and we can expect that to leading order the result for the area quantum does not depend on the dimension of the spacetime because the additional dimensions are ``hidden''. We believe that the problem is not straightforward, because the $(t,r)$ sector of the metric for the $d$-dimensional  Reissner-Nordstr\"om black hole depends on the dimension of the metric (for example see the factor $2 \mu / r^{d-3}$). Similar comments are valid for the $d$-dimensional Schwarzschild black hole.

In \cite{Setare:2003bd}, \cite{Barvinsky:2000gf,Medved:2009nj,Ropotenko:2009mh,Makela:1997rx} we find other results on the area spectrum of the Reissner-Nordstr\"om black hole (mainly in four dimensions). Setare \cite{Setare:2003bd} analyzes the four-dimensional extreme Reissner-Nordstr\"om black hole. Thus \cite{Setare:2003bd} and the present paper study different limits of the Reissner-Nordstr\"om solution and it is not possible to compare the obtained results for the area quantum. Also in \cite{Setare:2003bd} it is not shown that the area of the extreme Reissner-Nordstr\"om black hole behaves as an adiabatic invariant and it is expected that the extreme Reissner-Nordstr\"om black hole is a quantum object. Thus it is not straightforward that the Bekenstein and Hod ideas can be used to calculate the area quantum of the extreme Reissner-Nordstr\"om black hole.

In the small charge limit the area of the inner horizon is small and the total area is approximately the area of the event horizon, thus we may say that our results do not distinguish whether the total area or the event horizon surface is quantized; nevertheless we note that in the second method, to leading order we identify the area of the event horizon with the value of the integral for the adiabatic invariant $I$ (\ref{eq: adiabatic invariant RN}) of the Reissner-Nordstr\"om black hole (see also (\ref{e: Kwon Nam area I})) and hence we implicitly quantize the area of the outer horizon, and not the total area. 

Based on reduced phase-space quantization Barvinsky, et al.\ \cite{Barvinsky:2000gf} calculate the area spectrum of the $d$-dimensional Reissner-Nordstr\"om black hole (and other charged black holes). In \cite{Barvinsky:2000gf} it is quantized the quantity $A_B = A - A_0$, where $A$ is the area of the event horizon, $A_0$ is a function that depends on a positive power of the electric charge and is related to the horizon area of the extremal Reissner-Nordstr\"om black hole. In the small charge limit we can assume that $A_0$ is small and hence $A_B \approx A$. For the four-dimensional Reissner-Nordstr\"om black hole they find the two parameter area spectrum \cite{Barvinsky:2000gf} (a similar area spectrum is found for the $d$-dimensional Reissner-Nordstr\"om black hole)
\begin{equation} \label{eq: spectrum Barvinsky}
 A_{B} = 4 \pi \hbar (2 n + p +1), 
\end{equation} 
where $n,p=0,1,2,\dots$, with $p$ determining the charge spectrum $Q= \pm \sqrt{\hbar p}$. From formula (\ref{eq: spectrum Barvinsky}), if $Q$ is a constant, then $\Delta A_B = 8 \pi \hbar$, which is equal to our result. Thus in the small charge limit and for constant $Q$, our results to leading order are equal to those by Barvinsky, et al.\ \cite{Barvinsky:2000gf}. Moreover the area quantum that we get with the Kwon and Nam proposal is identical to $\Delta A_B$. In contrast to the method of Barvinsky, et al. \cite{Barvinsky:2000gf} the methods that we use in this paper do not determine the charge spectrum of the  Reissner-Nordstr\"om black hole.

A similar analysis to that of Barvinsky, et al.\ \cite{Barvinsky:2000gf} is developed by Ropotenko \cite{Ropotenko:2009mh} and Medved \cite{Medved:2009nj} for black holes whose metric can be written in Schwarzschild like form. In particular, for the area spectrum of the Reissner-Nordstr\"om black hole they find $A_n = 8 \pi \hbar n$. To leading order our result for the area quantum in the small charge limit  coincides with the value which is reported  in \cite{Medved:2009nj,Ropotenko:2009mh}, and the area spectrum that we get with the Kwon and Nam proposal is identical to that given in \cite{Medved:2009nj,Ropotenko:2009mh}.

\section{Summary}
\label{section 4}

From the results of \cite{Maggiore:2007nq}--\cite{Medved:2008iq} and from those of the previous sections, (for the $d$-dimensional Schwarzschild black hole, the $d$-dimensional Reissner-Nordstr\"om black hole in the small charge limit (to leading order), and the slowly rotating four-dimensional Kerr black hole) we obtain that their area quanta are equal to $\Delta A = 8 \pi \hbar$ and coincide with the results for $\Delta A$ already calculated \cite{Medved:2009nj}--\cite{Padmanabhan:2003ub}. At least to leading order, notice that this value of the area quantum is independent of the black hole parameters, the spacetime dimension, and the field parameters. We also note that for these three black holes, to leading order the two methods used in \cite{Maggiore:2007nq}--\cite{Medved:2008iq} and in the previous sections give the same value for the area quantum.

In contrast to original Hod's conjecture \cite{Hod:1998vk}, the previous results show that Hod's conjecture with the changes suggested by Maggiore \cite{Maggiore:2007nq} is applicable to black holes with two horizons, at least far from the extremal limit (see the examples of the slowly rotating four-dimensional Kerr black hole \cite{Vagenas:2008yi}, \cite{Medved:2008iq},  the  $d$-dimensional Reissner-Nordstr\"om black hole in the small charge limit to leading order and of the non extremal  Reissner-Nordstr\"om black hole). Thus for the black holes of Einstein-Maxwell theory, it is probable that for the dimensionless parameter $\epsilon$ of formula (\ref{eq: area spectrum}) the value $\epsilon = 8 \pi$ is generic (at least in some limit) as is suggested by Medved \cite{Medved:2004mh}.

Here for the non extremal Reissner-Nordstr\"om black holes we expand the analysis proposed by Kwon and Nam. Furthermore with the methods of Hod-Maggiore and Kunstatter-Maggiore we analyze the Reissner-Nordstr\"om black hole in the small charge limit. A question that deserves further research is to study whether the Hod-Maggiore and Kunstatter-Maggiore methods of the previous sections can be used when we eliminate this restriction on the mass and charge, but the black hole is still far from extremality. Notice that for this case in Kunstatter-Maggiore method the corrections of higher order must be considered.

\section{Acknowledgments}

This work was supported by CONACYT M\'exico, SNI M\'exico, EDI-IPN, COFAA-IPN, and Research Projects SIP-20100777 and SIP-20100684.


\end{document}